\documentclass[pdflatex,sn-mathphys-num]{sn-jnl}
\usepackage{graphicx}%
\usepackage{multirow}%
\usepackage{amsmath,amssymb,amsfonts}%
\usepackage{amsthm}%
\usepackage{mathrsfs}%
\usepackage[title]{appendix}%
\usepackage{xcolor}%
\usepackage{textcomp}%
\usepackage{manyfoot}%
\usepackage{booktabs}%
\usepackage{array}%
\usepackage{algorithm}%
\usepackage{algorithmicx}%
\usepackage{algpseudocode}%
\usepackage{listings}%
\usepackage{caption}
\usepackage{tikz}
\usepackage{tikz-3dplot}
\usetikzlibrary{arrows.meta, calc}
\usepackage{hhline}

\theoremstyle{thmstyleone}%

\theoremstyle{thmstyletwo}%

\theoremstyle{thmstylethree}%

\begin{document}

\title[Article Title]{Emergence of Strategic Equilibria from Transverse Field Ising Hamiltonian Dynamics}

\author[1]{\fnm{Teena} \sur{Thomas}}\email{teena.thomas@vit.ac.in}

\author[1]{\fnm{S.} \sur{Balakrishnan}}\email{physicsbalki@gmail.com}

\affil[1]{\orgdiv{Department of Physics, School of Advanced Sciences}, \orgname{Vellore Institute of Technology}, \orgaddress{\street{Vellore}, \postcode{632014}, \state{Tamil Nadu}, \country{India}}}

\abstract{Game theory studies strategic decision-making among rational agents, and many classical games can be mapped onto interaction models such as the Ising model. Quantum game theory extends this framework by allowing players to exploit quantum superposition and entanglement. In this work, we study quantum games using an operator-based formulation derived from the transverse-field quantum Ising model. We show that the Hamiltonian-driven dynamics naturally generate entangling operator which resolve the dilemma in the game settings. This leads to a clear quantum advantage over classical outcomes. Unlike standard quantization schemes based on fixed entangling gates, the present approach enables tunable entanglement, controlled directly by physical Hamiltonian parameters, providing a hardware-relevant perspective on quantum game design.
}

\keywords{Quantum Game theory, Quantum Ising model, Quantum statistical physics, entanglement, operator, modified EWL scheme}

\maketitle
\section{Introduction}\label{sec1}
Game theory provides a formal mathematical framework for analyzing situations in which multiple decision-makers interact strategically, with each participant’s outcome depending on the choices made by others \cite{sanz2025mapping}. Since its foundational development \cite{von1992theory}, game-theoretic framework has been used to model a wide variety of real-world scenario, ranging from economic competition and social behavior to biological and political systems. In such models, each “player” selects an available strategy with the objective of optimizing an associated payoff, which may represent profit, survival, or overall success. A diverse set of games including the prisoner’s dilemma, battle of the sexes \cite{marinatto2000quantum}, chicken game \cite{iqbal2001evolutionarily}, stag hunt \cite{li2012quantum}, matching pennies \cite{balakrishnan2013classical}, and public goods game \cite{benjamin2019triggers} has been extensively investigated to capture different patterns of strategic interaction \cite{iqbal2005studies}.\\
A notable connection exists between game-theoretic models and concepts from statistical physics, particularly those associated with the Ising model \cite{galam2010ising}. This analogy arises from the payoff maximization in games and energy minimization in physical systems. While players adjust their strategies to achieve optimal outcomes say, spins in a magnetic system align so as to minimize the system’s energy. As a result, various strategic games can be mapped onto Ising-type Hamiltonians, where equilibrium states correspond to stable strategic configurations. In large populations, tools from statistical mechanics become especially useful, as they allow one to study collective behavior such as the emergence or breakdown of cooperation in the game setting.\\
Traditional game theory restricts players to classical, well-defined strategies. This limitation was addressed with the development of quantum game theory \cite{wiesner1983conjugate}, \cite{meyer1999quantum}, which extends the classical framework by permitting strategies that exploit quantum mechanical principles \cite{grabbe2005introduction}. In this setting, players operate on quantum states using unitary transformations, enabling strategic choices that can exist in superposition and become entangled \cite{anand2015quantum}, \cite{vyas2021essence}, \cite{sarkar2019quantum}. By incorporating quantum resources, quantum game theory offers a richer strategic space and reveals behaviors that have no classical counterpart \cite{flitney2002introduction}, \cite{piotrowski2003invitation}, \cite{hidalgo2008quantum}, \cite{guo2008survey}. Consequently, when many quantum players interact, the resulting system cannot be adequately described using classical statistics, instead requires the tools of quantum statistical mechanics to characterize the corresponding ensemble \cite{benjamin2025agent}.\\
Despite significant progress in quantum game theory, most existing quantization schemes rely on gate-based constructions, such as the Eisert-Wilkens-Lewenstein (EWL) protocol \cite{eisert1999quantum}, where entanglement is introduced through a fixed entangling gate. In contrast, the transverse-field Ising model naturally incorporates both interaction-induced correlations and quantum fluctuations through its competing coupling and field terms. This makes it an ideal candidate for studying how quantum strategies emerge from physically motivated Hamiltonian dynamics. In this work, we adopt a Hamiltonian-driven formulation of quantum games based on the transverse-field Ising model which allows us to identify tunable entangling dynamics and examine their impact on equilibrium behavior beyond classical and standard quantum game formulations.\\
The paper is organized as follows. Section \ref{sec2} introduces the transverse field quantum Ising model that serves as the physical framework for our analysis. Within this section, we first recall the classical Ising model and its game-theoretic interpretation, and then extend the discussion to the quantum regime. The non-interacting limit is examined to clarify the emergence of trivial strategic behavior, followed by the construction of an operator-based quantum game-theoretic framework derived from hamiltonian time evolution. In Section \ref{sec3}, we employ a modified Eisert–Wilkens–Lewenstein quantization scheme to realize quantum games through transverse field Ising model-induced entanglement, and investigate how the resulting equilibria depend on tunable Hamiltonian parameters. Section \ref{sec4} presents a detailed equilibrium analysis of specific quantum games, namely the Prisoner’s Dilemma, the Game of Chicken, and the Stag Hunt, discussed in separate subsections. Finally, Section \ref{sec5} concludes with a discussion of the main results and their implications.
\section{Transverse Field Quantum Ising Model}\label{sec2}
In this section, we adopt the transverse field Ising model (TFIM) as the physical framework for our analysis. This model describes interacting spin systems with nearest-neighbor Ising coupling of strength $J$ and a transverse magnetic field that introduces non-commuting spin dynamics. Before proceeding to the quantum regimes, it is instructive to recall the classical Ising model and its established correspondence with game theory.
\subsection{Classical Ising Model and its Game-Theoretic Interpretation}\label{sec2.1}
There exists a well-established correspondence between the classical game theory and the Ising model \cite{galam2010ising}. Each lattice site,  occupied by an atomic spin, can be viewed as a player and the spin states $+1$ and $-1$ represent the available strategies. Interactions between neighboring spins encode strategic interdependence, and the system’s tendency to minimize its total energy shows the players’ objective of maximizing individual payoffs.\\
In reference \cite{tejasvi2022study}, it is clearly shown that the classical Ising model can be used to study Potential games. In contrast to purely competitive settings, a potential game describes a situation in which players are effectively stakeholders in a common objective. Rather than optimizing independent payoff functions, all players’ incentives are aligned with a single global quantity, known as the potential function.\\
Within this mapping, the system magnetization emerges as a key observable. Magnetization measures the net alignment of spins and therefore quantifies the collective strategic preference of the population toward one strategy over the other. In the thermodynamic limit, magnetization provides insight into equilibrium selection, the stability of strategic outcomes, and the prevalence of cooperation or defection throughout the population \cite{adami2018thermodynamics}, \cite{benjamin2020thermodynamic}. This classical correspondence motivates the extension to quantum spin systems, where non-commuting operators and entangling dynamics introduce quantum strategic effects.
\subsection{Quantum Ising Model}\label{sec2.2}
\begin{figure*}[t]
\centering
\includegraphics[width=\textwidth]{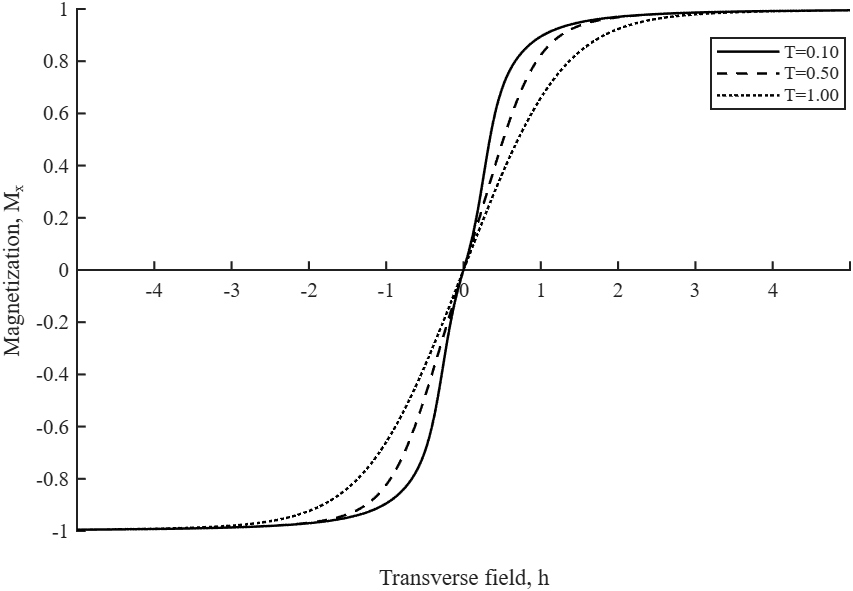}
\caption{Magnetization versus transverse field at different temperatures.}
\label{fig:temp}
\end{figure*}

We now introduce the transverse field quantum Ising model that governs the quantum dynamics of the spin system considered in this work. In quantum mechanics, spins are represented by pauli operators $(\sigma^x, \sigma^y, \sigma^z)$. In the case of the Transverse Field Quantum Ising Model, spins are considered to be in either $+z$ or $-z$ direction and external magnetic field is introduced in the transverse direction ($x$~axis) which can align the spins in the respecive field direction. Pauli-$x$ operator ($\sigma^x$) that do not commute with the pauli-$z$ operator ($\sigma^z$) introduce quantum fluctuations.\\
Its Hamiltonian is given by \cite{rose}, \cite{sachdev1999quantum},
\begin{equation}
    H=-J \sum\limits_{<i,j>} \sigma_i^{z} \sigma_j^{z} - h \sum\limits_i \sigma_i^{x}
    \label{eq:ham}
\end{equation}
where $\sigma^x$ is the Pauli-$x$ operator, ${\sigma_i}^z {\sigma_j}^z$ represents the Pauli-$z$ operator acting on the $i^{th}$ and $j^{th}$ spins respectively. $J$ denotes the coupling constant that characterizes the interaction among the spins that are directed in the z direction, $h$ represents the transverse magnetic field. The second term, $h\sigma^x$, tries to flip spins, introducing the superposition between spin states.\\
Magnetization along the transverse direction (for the case of two spins) is found using the equation,
\begin{equation}
    M_x= Tr(\rho ~(\sigma^x\otimes I+I \otimes\sigma^x))
\end{equation}
where $Tr$, $\rho$, $I$ represent the trace of the matrix, the density matrix and  the Identity operator respectively.\\
On substituting the values, magnetization expression in terms of the Ising parameter is derived as,
\begin{equation}
    M_x= \frac{2h~\sinh(\beta \sqrt{J^2+4h^2})} {\sqrt{J^2+4h^2}(\cosh(\beta J)+\cosh(\beta \sqrt{J^2+4h^2}))}
\end{equation}
where $\beta=1/kT$ is the inverse temperature and $k$ denotes the Boltzmann constant.\\
$M_x$ against the transverse field $h$ for different temperatures at a fixed interaction strength ($J=1$) is plotted in Figure \ref{fig:temp}.
The following convention has been taken for the analysis:\\
A value of	$M_x=+1$ and $M_x=-1$ indicate alignment of both the spins along the positive $x$-direction and alignment of both the spins along the negative $x$ direction respectively whereas $M_x = 0$ shows that both the spins are not aligned with field.\\
In game point of view, magnetization is the difference in number of players choosing one strategy over the other. In classical Ising model, magnetization along z direction ($M_z$) is considered in which $M_z=+1$ shows that all players are adopting the $Cooperate$ strategy (all spins up), while $M_z=-1$ represents that all players are adopting the $Defect$ strategy 
(all spins down). Since external control is applied along the transverse direction, transverse magnetization $M_x$ is focused here, directly captures the system’s response to the applied field and reflects the influence of quantum fluctuations. Thus, $M_x=+1$ shows that all players are adopting the $Cooperate$ strategy (both spins aligned in $+x$ axis) whereas $M_x=-1$ represents that all players are adopting $Defect$ strategy (both spins aligned in $-x$ axis). Thus, transverse magnetic field acts as an external incentive that biases players toward a particular strategic direction. From Figure \ref{fig:temp}, it is clear that as the strength of magnetic field in a particular transverse direction increases, the possibility of all spins being aligned in that particular direction increases which can be interpreted as the possibility of players to choose one particular strategy depending on the external incentives. Also, the temperature may be viewed as a source of uncertainty in strategic decision-making. At low temperatures, where fluctuations are minimal, even a weak incentive produces a sharp collective response, leading to rapid alignment. In contrast, at higher temperatures, increased uncertainty weakens coordinated behavior, requiring stronger incentives to induce a dominant strategic tendency. That is, higher uncertainty suppresses coordinated behavior, necessitating stronger incentives to produce a dominant strategy. A similar behavior is obtained in the classical Ising model \cite{benjamin2020emergence}.\\
\subsection{Non-Interacting Limit and the Emergence of a Lonely Game}\label{sec2.3}
For exploring the Ising pay-off matrix, consider the case of two players. In that case, the Hamiltonian reduces to,
\begin{equation}
   H=-J \sigma_1^{z} \sigma_2^{z} - h (\sigma_1^{x}+\sigma_2^{x})\\
\end{equation}
We first consider the non-interacting limit 
$J=0$, where no coupling exists between the two spins. In this regime, entanglement is absent and the Hamiltonian simplifies to
\begin{equation}
    H=- h (\sigma_1^{x}+\sigma_2^{x})
\end{equation}
The individual energy contributions associated with spins $1$ and $2$ are then given by
\begin{equation}
\centerline{$E_1=-h\sigma_1^{x},~E_2=-h\sigma_2^{x}$}
\end{equation}
$|+x\rangle$ and $|-x\rangle$ are the eigen states of $\sigma^{x}$ with eigen values $+1$ and $-1$ respectively. Taking the eigen states as the strategies, the quantum Ising payoff matrix for the two players can be written as\\
\begin{equation}
\left(
\begin{array}{c|c c}
 & |+x\rangle & |-x\rangle \\ \hline
 |+x\rangle & (+h,+h) & (+h,-h) \\
 |-x\rangle & (-h,+h) & (-h,-h)
\end{array}
\right)
\label{eq:payoff_sigma}
\end{equation}
The above matrix resembles the payoff matrix of the game called 'lonely game' and is studied in \cite{galam2010ising}.\\
This matrix is formally analogous to the payoff structure of the so-called lonely game. In such a game, each player makes decisions independently of the other’s choice, and the resulting dominant-strategy Nash equilibrium fails to be Pareto optimal. For instance, a player may independently decide whether to undertake a particular action, such as purchasing a toy, receiving a positive payoff if the action is taken and a negative payoff otherwise, regardless of the other player’s behavior \cite{new}.
\subsection{Operator-Based Quantum Game Theoretic Framework}\label{sec2.4}
\begin{figure*}[h]
 \centering
  \includegraphics[width=\textwidth]{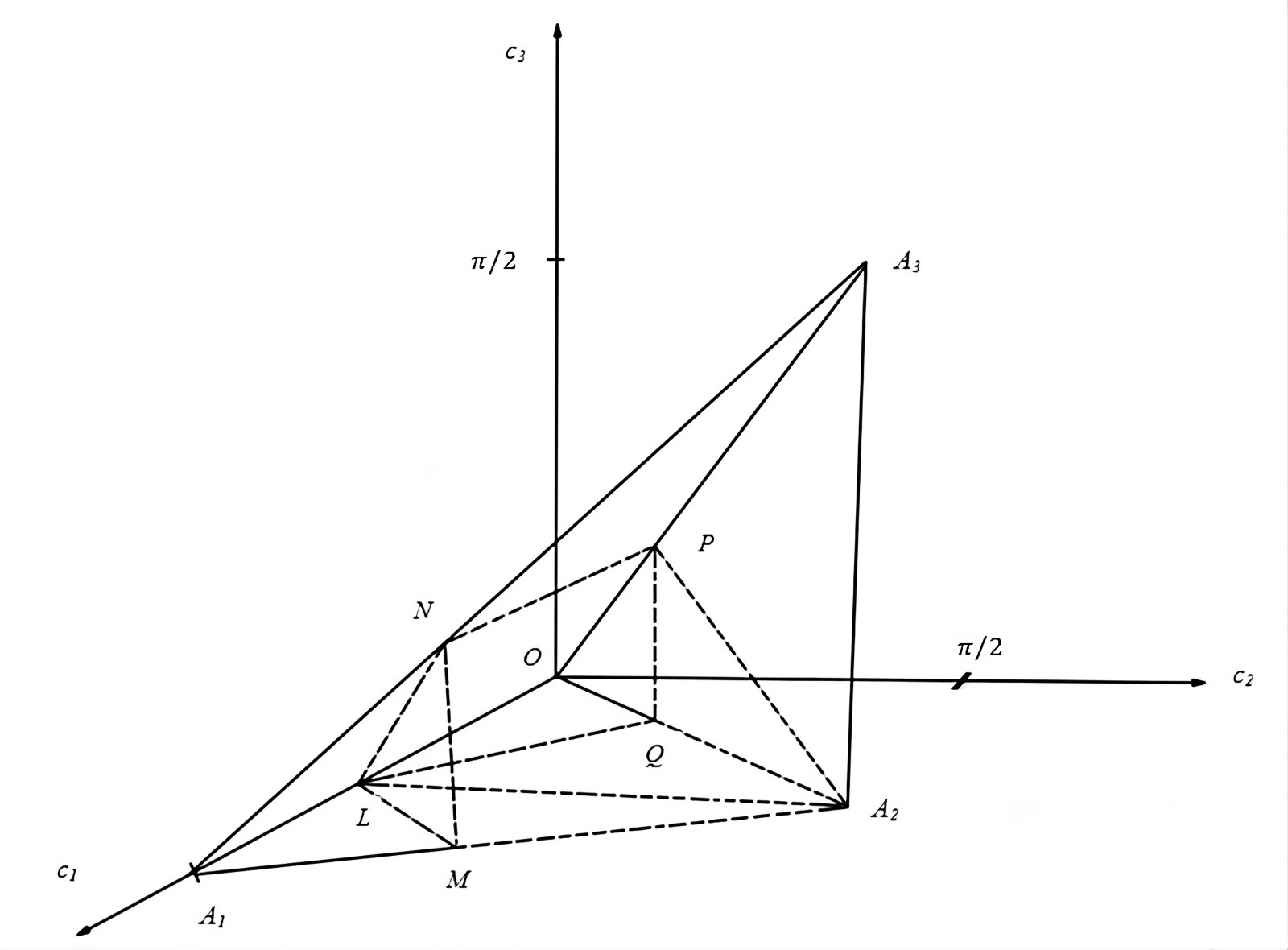}
 \caption{Weyl chamber: Geometrical representation of two-qubit operators.}
 \label{fig:3}
\end{figure*}

\begin{figure*}[htbp]
\centering
\includegraphics[width=0.7\textwidth]{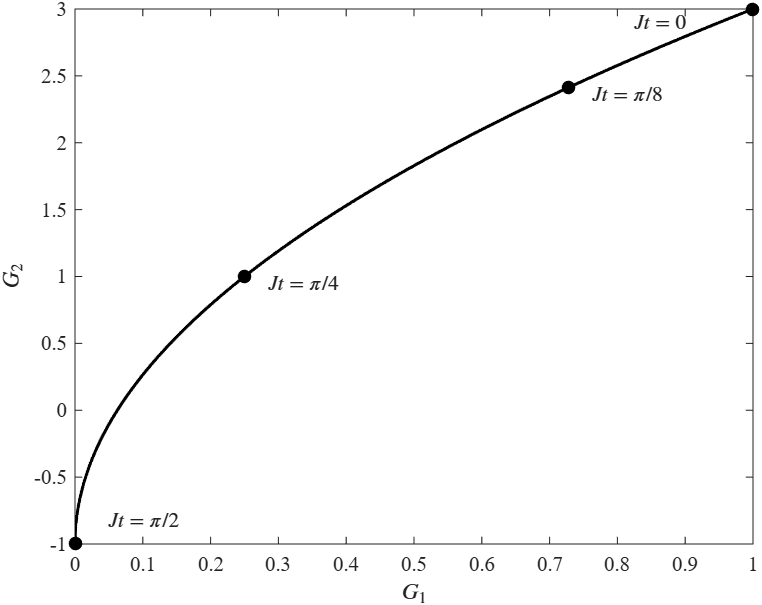}
\caption{Parametric plot of invariants with angle labels.}
\label{fig:invariants}
\end{figure*}

\begin{figure*}[htbp]
\centering
\includegraphics[width=0.7\textwidth]{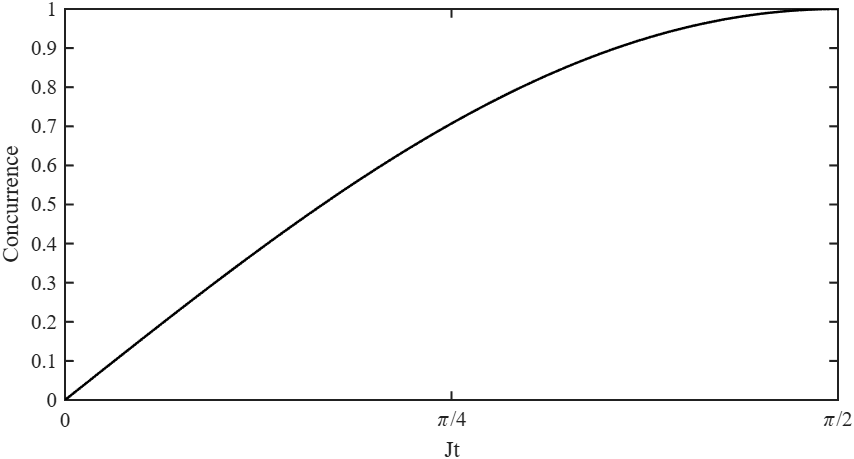}
\caption{Variation of concurrence with $Jt$.\\For $Jt = 0$, $\pi/4$, $\pi/2$, concurrence = $0, 0.707, 1$ respectively.}
\label{fig:concurrence}
\end{figure*}

When $J\neq0$, the payoff matrix can no longer be expressed in the form given in \eqref{eq:payoff_sigma}. This is because, unlike in the classical Ising model, transverse quantum Ising model has the pauli-$x$ as well as pauli-$z$ operators. To analyze the quantum games that can be realized using the transverse field quantum Ising model, we explored the operator part. For that, corresponding unitary operator is found out using,
\begin{equation}
U=e^{-iHt}
\end{equation}
where $H$ is the Hamiltonian corresponding to transverse field quantum Ising model and $t$ is the time evolved.\\
With the eigenvectors and eigenvalues,
\begin{equation}
U = V e^{-iFt} V^\dagger
\end{equation}
where $V$ is the eigenvector matrix and $F$ is the diagonalized matrix. This results in a $4$x$4$ unitary operator,\\
\begin{equation}
\begin{pmatrix}
a_{11} & a_{12} & a_{13} & a_{14} \\
a_{21} & a_{22} & a_{23} & a_{24}\\
a_{31} & a_{32} & a_{33} & a_{34} \\
a_{41} & a_{42} & a_{43} & a_{44}
\label{eq:U}
\end{pmatrix}
\end{equation}
Thus, we got a set of $16$ elements in terms of $J$, $h$ and $t$ and expressed as
\begin{equation}
\begin{aligned}
&a_{11}=a_{44}=e^{iJt/2}+\frac{h^{2} e^{-i\sqrt{J^{2}+4h^{2}}\,t}}{J^{2}+4h^{2}+J\sqrt{J^{2}+4h^{2}}} +\frac{h^{2}e^{i\sqrt{J^{2}+4h^{2}}\,t}}{J^{2}+4h^{2}-J\sqrt{J^{2}+4h^{2}}} 
\label{eq:1}
\end{aligned}
\end{equation}
\begin{equation}
\begin{aligned}
a_{12}= a_{13}=a_{21}=a_{24}=a_{31}=a_{34}=a_{42}=a_{43}
&=\small \frac{h(-J-\sqrt{J^2+4h^2}) e^{-i\sqrt{J^{2}+4h^{2}}\,t}}{2\left(J^{2}+4h^{2}+J\sqrt{J^{2}+4h^{2}}\right)} \\&\quad+ \frac{h(-J+\sqrt{J^2+4h^2}) e^{i\sqrt{J^{2}+4h^{2}}\,t}}{2\left(J^{2}+4h^{2}-J\sqrt{J^{2}+4h^{2}}\right)}
\label{eq:2}
\end{aligned}
\end{equation}
\begin{equation}
\begin{aligned}
&a_{14}=a_{41}=-e^{iJt/2}+   \frac{h^{2} e^{-i\sqrt{J^{2}+4h^{2}}\,t}}{J^{2}+4h^{2}+J\sqrt{J^{2}+4h^{2}}} +\frac{h^{2}e^{i\sqrt{J^{2}+4h^{2}}\,t}}{J^{2}+4h^{2}-J\sqrt{J^{2}+4h^{2}}} 
\label{eq:3}
\end{aligned}
\end{equation}
\begin{equation}
\begin{aligned}
a_{22}=a_{33}=e^{-iJt/2}+\frac{(-J-\sqrt{J^2+4h^2})^{2} e^{-i\sqrt{J^{2}+4h^{2}}\,t}}{4(J^{2}+4h^{2}+J\sqrt{J^{2}+4h^{2})}} + \frac{(-J+\sqrt{J^2+4h^2})^{2} e^{i\sqrt{J^{2}+4h^{2}}\,t}}{4(J^{2}+4h^{2}-J\sqrt{J^{2}+4h^{2})}} 
\label{eq:4}
\end{aligned}
\end{equation}
\begin{equation}
\small
\begin{aligned}
a_{23}=a_{32}=-e^{-iJt/2}+\frac{(-J-\sqrt{J^2+4h^2})^{2} e^{-i\sqrt{J^{2}+4h^{2}}\,t}}{4(J^{2}+4h^{2}+J\sqrt{J^{2}+4h^{2})}} + \frac{(-J+\sqrt{J^2+4h^2})^{2} e^{i\sqrt{J^{2}+4h^{2}}\,t}}{4(J^{2}+4h^{2}-J\sqrt{J^{2}+4h^{2})}} 
\label{eq:5}
\end{aligned}
\end{equation}
This is compared with the general form of two-qubit gate, expressed using geometrical points ($c_1, c_2, c_3$).\\ 
\begin{equation}
U \equiv
\begin{pmatrix}
e^{-\frac{i c_{3}}{2}}\, c^{-} & 0 & 0 & -i e^{-\frac{i c_{3}}{2}}\, s^{-} \\
0 & e^{\frac{i c_{3}}{2}}\, c^{+} & -i e^{\frac{i c_{3}}{2}}\, s^{+} & 0 \\
0 & -i e^{\frac{i c_{3}}{2}}\, s^{+} & e^{\frac{i c_{3}}{2}}\, c^{+} & 0 \\
- i e^{-\frac{i c_{3}}{2}}\, s^{-} & 0 & 0 & e^{-\frac{i c_{3}}{2}}\, c^{-}
\label{eq:u}
\end{pmatrix}
\end{equation}
where
$c^{\pm} = \cos\!\left(\frac{c_{1} \pm c_{2}}{2}\right), 
\qquad 
s^{\pm} = \sin\!\left(\frac{c_{1} \pm c_{2}}{2}\right).$\\
By solving element-wise from the matrices (Eq. \eqref{eq:U} and Eq. \eqref{eq:u}), we obtain,
\begin{equation}
\label{eq:1c}
(c_1, c_2, c_3)
=\left(0,-\pi - \frac{J\pi}{\sqrt{J^2 + 4h^2}},\pi -\frac{J\pi}{\sqrt{J^2 + 4h^2}}
\right)
\end{equation}
and
\begin{equation}
    t=\frac{\pi}{\sqrt{J^2+4h^2}}
\end{equation}
This allows us to express the geometrical points in terms of the physical parameters $J$,$h$ and $t$. However, due to local-unitary equivalence, this representation is not unique and does not necessarily lie within the canonical Weyl chamber, the geometrical representation of two-qubit operators (shown in Fig. \ref{fig:3}). Shifts of the form $c_k~\rightarrow~c_k+\pi$ leave the non-local content invariant, as they can be absorbed into local operations. Applying these modulo-$\pi$ reductions and exploiting Weyl group symmetries like sign changes and permutations, the parameter set is mapped into the canonical Weyl chamber. 
\begin{equation}
 \centerline{$(c_1,c_2,c_3)=(Jt,Jt,0)$ }   
\end{equation}
which lies along the $OA_2$ edge of the Weyl chamber.
Along this edge, the physically distinct nonlocal gates are parameterized by,
$0\le\theta=Jt\le\pi/2$, yielding a family of gates in which $Jt=0$, $Jt=\pi/4$, $Jt=\pi/2$  correspond to Cartan coordinates that are locally equivalent to the $Identity$ ($0,0,0$), $\sqrt{iSWAP}$ ($\pi/4,\pi/4,0$) and $iSWAP$ ($\pi/2,\pi/2,0$) gates, respectively.\\
Within the Weyl chamber, perfect entanglers occupy a well-defined sub-region characterized by simple inequalities among the Cartan parameters. A two-qubit gate belongs to the class of perfect entanglers, if its Cartan coordinates satisfy the inequalities, $c_1+c_2\ge\pi/2$ and $c_2+c_3\le\pi/2$. For the present parameterization along the $OA_2$ edge, these conditions reduce to,
\begin{equation*}
    \pi/4\le Jt \le \pi/2.
\end{equation*}
To further confirm the entangling nature independent of local operations, the local invariants $G_1$ and $G_2$ are found out from \cite{rezakhani2004characterization} as,\\
$G_1=\frac{1}{4}[e^{-ic_3}cos(c_1-c_2)+e^{ic_3}cos(c_1+c_2)]^2$ and $G_2=cos~2c_1+cos~2c_2+cos~2c_3$ as
\begin{equation}
\centerline{
  $G_1=\frac{1}{4}[1+cos~2Jt]^2$,
\qquad
$G_2=1+2~cos~2Jt$.}
\end{equation}
$G_1$ and $G_2$ are plotted for different values of $Jt$ in Fig.~\ref{fig:invariants}.
Unlike the standard EWL quantization scheme, which relies on a fixed maximally entangling $CNOT$ gate, the TFIM naturally generates a continuum of locally inequivalent entangling gates, including a complete family of perfect entanglers along the $QA_2$ edge (see Fig. \ref{fig:3}).\\
To further characterize the entangling capability of the generated gates, the concurrence of the evolved two-qubit state is evaluated. To characterize the evolution, the TFIM generated cartan parameters are substituted in \eqref{eq:u} and the general superposition state is considered as the initial state $|\psi_0\rangle$.
\begin{equation*}
   |\psi_0\rangle=\frac{1}{2}(|00\rangle+|01\rangle+|10\rangle+|11\rangle) 
\end{equation*}
The expression for concurrence is obtained as,
\begin{equation}
    Concurrence= 2 |\frac{1}{4}(1-e^{-i2Jt})|
    \end{equation}
and plotted in Fig.~\ref{fig:concurrence}.
The concurrence increases monotonically with the parameter $Jt$, implying the progressive generation of entanglement under TFIM evolution. At $Jt=0$, the system remains in a separable state corresponding to the Identity operation, yielding zero concurrence. As $Jt$ increases to $\pi/4$, the system evolves into a partially entangled state associated with the $\sqrt{iSWAP}$ gate with concurrence reaching $\frac{1}{\sqrt{2}}$. Finally, at $Jt=\pi/2$, the evolution corresponds to the $iSWAP$ gate, producing a maximally entangled state with unit concurrence. This behavior highlights that the parameter $Jt$ serves as an effective and continuous control for tuning entanglement in the system. Thus, the TFIM provides a physically controllable pathway to realize entangling gates.
\section{Modified EWL Scheme for TFIM Hamiltonian-Driven Games}\label{sec3}
The pay-off matrix for a player in the case of a two-player game is
\begin{equation}
p =
\begin{pmatrix}
p_{00} & p_{01} \\
p_{10} & p_{11}
\end{pmatrix}
\label{eq:payoffmatrix}
\end{equation}
The payoff matrices for any two- player game can be represented in the most general form as
\begin{equation}
p_{player
_1} =
\begin{pmatrix}
X & Y \\
W & Z
\label{eq:payoffmatrixalice}
\end{pmatrix},
\qquad
p_{player_2}=
\begin{pmatrix}
R & S \\
B & V
\end{pmatrix}
\end{equation}
According to the EWL scheme, the final state can be represented as,
\begin{equation}
|\psi_f\rangle = U^{\dagger} (k_1 \otimes k_2) U |\psi_i\rangle
\end{equation}
where $|\psi_i\rangle$ refers to the initial state, $U$ is the two-qubit entangling operator, $k_1 \otimes k_2$ are the unitary operations which corresponds to the strategies performed by the players, and $U^{\dagger}$ is the disentangling operator. A projective measurement is performed on the computational basis on the final state and using \eqref{eq:payoffmatrix}, expectation value (average pay-off) of the players can be calculated as,
\begin{equation}
\langle \$ \rangle
= p_{00} |\langle \psi_f | 00 \rangle|^2
+ p_{01} |\langle \psi_f | 01 \rangle|^2
+ p_{10} |\langle \psi_f | 10 \rangle|^2
+ p_{11} |\langle \psi_f | 11 \rangle|^2
\end{equation}
Ref.~\cite{vijayakrishnan2019role} introduced a modified EWL scheme and evaluated the average pay-offs of $player_1$ and $player_2$ in case of pure strategies. In the present work, we follow the same scheme and for our analysis, we choose the initial state as $| 00 \rangle$ and compute the average payoffs of the two players for different strategy combinations. The resulting payoffs are summarized in Table~\ref{tab:payoff}.
\begin{table}[htbp]
\small
\centering
\caption{Average payoffs of $player_1$ and $player_2$ under different pure strategies.}
\label{tab:payoff}
\begin{tabular}{lcc}
\toprule
Strategies & $\langle \$ \rangle_{player_1}$ & $\langle \$ \rangle_{player_2}$ \\
\midrule
$I \otimes I$ 
& $X$ 
& $R$ \\

$I \otimes \sigma^x$ 
& $Y \cos^{2} c_{2} + W \sin^{2} c_{2}$ 
& $S \cos^{2} c_{2} + B \sin^{2} c_{2}$ \\

$\sigma^x \otimes I$ 
& $Y \sin^{2} c_{2} + W \cos^{2} c_{2}$ 
& $S \sin^{2} c_{2} + B \cos^{2} c_{2}$ \\

$\sigma^x \otimes \sigma^x$ 
& $Z$ 
& $V$ \\
\bottomrule
\end{tabular}
\end{table}\\
$I$ and $\sigma^x$ corresponds to cooperation and defection strategies respectively.\\
For the TFIM-generated evolution considered here, the Cartan parameter $c_2$ is given by $c_2 = Jt$ and on substituting the value, we get,
\begin{equation}
\small
p_{player_1}=
 \left(
\begin{array}{c|c c}
 & I & \sigma^x \\ \hline
 I & (X,R) & (Y\,\cos^2{Jt}+ W\,\sin^2{Jt},\; S\,\cos^2{Jt}+ B\,\sin^2{Jt}) \\
 \sigma^x & (Y\,\sin^2{Jt}+W\,\cos^2{Jt},\; S\,\sin^2{Jt}+B\,\cos^2{Jt}) & (Z,V)
 \end{array}
 \right)
\label{eq:payoffJt}
\end{equation}
where the first elements show payoff of $player_1$ whereas second elements show that of $player_2$. 
\section{Analysis of Equilibrium in Quantum Games}\label{sec4}
\subsection{Prisoner's Dilemma}
Consider the general form of payoff matrix used for prisoner's dilemma,
\begin{equation}
p_1=
 \left(
\begin{array}{c|c c}
 &Cooperate(C) & Defect (D) \\ \hline
 Cooperate(C) & (3,3) & (0,5)\\
 Defect(D) &(5,0) &    (1,1)
 \end{array}
 \right)   
\end{equation}
Comparing this equation with Eq.~\eqref{eq:payoffJt} and substituting different values of $Jt$ like $0$, $\pi/4$, $\pi/2$, we get,
\begin{equation}
p_1^{'} =
\begin{array}{c c c}
\left(
\begin{array}{c|cc}
    & C & D \\
\hline
C & (3,3) & (0,5) \\
D & (5,0) & (1,1)
\end{array}
\right),
&
\left(
\begin{array}{c|cc}
    & C & D \\
\hline
C & (3,3) & \left (\tfrac{5}{2},\tfrac{5}{2}\right) \\
D & \left (\tfrac{5}{2},\tfrac{5}{2}\right) & (1,1)
\end{array}
\right),
&
\left(
\begin{array}{c|cc}
    & C & D \\
\hline
C & (3,3) & (5,0) \\
D & (0,5) & (1,1)
\end{array}
\right)
\\[6pt]
\text{$Jt=0$}
&
\text{$Jt=\pi/4$}
&
\text{$Jt=\pi/2$}
\end{array}
\end{equation}
In the classical limit $Jt=0$, corresponding to a vanishing nonlocal interaction and a gate that is locally equivalent to the $Identity$, the Prisoner’s Dilemma exhibits the well-known Nash equilibrium ($D,D$), where both players defect despite mutual cooperation yielding a higher joint payoff. This outcome represents the fundamental dilemma of the game, in which rational decision-making leads to a socially sub-optimal result. As the entangling interaction is increased to $Jt=\pi/4$, placing the system in an entanglement regime equivalent to the $\sqrt{iSWAP}$ gate class, the Nash equilibrium shifts to ($C,C$), indicating that mutual cooperation becomes stable against unilateral deviation. Importantly, this transition occurs without modifying the payoff matrix or the available strategies, and is driven solely by quantum entanglement. For maximal entanglement at $Jt=\pi/2$, which is locally equivalent to the $iSWAP$ gate, the cooperative equilibrium ($C,C$) persists, demonstrating that entanglement completely resolves the dilemma by stabilizing cooperation as the unique rational outcome.
\subsection{Game of Chicken}
Consider the general form of payoff matrix used for game of chicken,
\begin{equation}
p_2=
 \left(
\begin{array}{c|c c}
 &Cooperate(C) & Defect(D) \\ \hline
 Cooperate(C) & (3,3) & (1,5)\\
 Defect(D) &(5,1) &    (0,0)
 \end{array}
 \right)   
\end{equation}
Comparing this equation with Eq.~\eqref{eq:payoffJt} and substituting different values of $Jt$ like 0, $\pi/4$, $\pi/2$, we get,
\begin{equation}
p_2^{'} =
\begin{array}{c c c}
\left(
\begin{array}{c|cc}
    & C & D \\
\hline
C & (3,3) & (1,5) \\
D & (5,1) & (0,0)
\end{array}
\right),
&
\left(
\begin{array}{c|cc}
    & C & D \\
\hline
C & (3,3) & (3,3) \\
D & (3,3) & (0,0)
\end{array}
\right),
&
\left(
\begin{array}{c|cc}
    & C & D \\
\hline
C & (3,3) & (5,1) \\
D & (1,5) & (0,0)
\end{array}
\right)
\\[6pt]
\text{$Jt=0$}
&
\text{$Jt=\pi/4$}
&
\text{$Jt=\pi/2$}
\end{array}
\end{equation}
For the Game of Chicken in the classical regime $Jt=0$, an evolution locally equivalent to the $Identity$, yields two asymmetric Nash equilibria, ($C,D$) and ($D,C$), reflecting the intrinsic anti-coordination nature of the game. In this scenario, each player benefits by defecting while the other cooperates, leading to strategic conflict and the absence of a fair symmetric outcome. When the entanglement is increased to $Jt=\pi/4$, placing the system in an entanglement regime locally equivalent to the $\sqrt{iSWAP}$ gate class, the equilibrium structure is enriched, and the symmetric cooperative outcome ($C,C$) emerges in addition to the classical asymmetric equilibria. For stronger interactions within the perfect-entangler region including $Jt=\pi/2$, which is locally equivalent to the $iSWAP$ gate class, the asymmetric equilibria disappear and ($C,C$) becomes the unique Nash equilibrium. Thus, entanglement not only promotes cooperation but also eliminates inequitable outcomes, steering the game toward a fair and risk-free resolution.
\subsection{Stag Hunt Game}
Consider the general form of payoff matrix used for stag hunt game,
\begin{equation}
p_3=
 \left(
\begin{array}{c|c c}
 &Cooperate(C) & Defect(D) \\ \hline
 Cooperate(C) & (4,4) & (0,3)\\
 Defect(D) &(3,0) &    (2,2)
 \end{array}
 \right)   
\end{equation}
Comparing this equation with Eq.~\eqref{eq:payoffJt} and substituting different values of $Jt$ like 0, $\pi/4$, $\pi/2$. The respective payoff matrices are,

\begin{equation}
p_3^{'} =
\begin{array}{c c c}
\left(
\begin{array}{c|cc}
    & C & D \\
\hline
C & (4,4) & (0,3) \\
D & (3,0) & (2,2)
\end{array}
\right),
&
\left(
\begin{array}{c|cc}
    & C & D \\
\hline
C & (4,4) & \left(\tfrac{3}{2},\tfrac{3}{2}\right) \\
D & \left(\tfrac{3}{2},\tfrac{3}{2}\right) & (2,2)
\end{array}
\right),
&
\left(
\begin{array}{c|cc}
    & C & D \\
\hline
C & (4,4) & (3,0) \\
D & (0,3) & (2,2)
\end{array}
\right)
\\[6pt]
\text{$Jt=0$}
&
\text{$Jt=\pi/4$}
&
\text{$Jt=\pi/2$}
\end{array}
\end{equation}
In the classical Stag Hunt game at $Jt=0$, corresponding to the absence of nonlocal interactions and an evolution locally equivalent to  $Identity$ results in two Nash equilibria, ($C,C$) and ($D,D$), representing a coordination problem in which cooperation is payoff-dominant but risky, while defection offers a safer but inferior outcome. The presence of multiple equilibria can lead to coordination failure, as players may rationally choose defection to avoid uncertainty. As the interaction strength increases to $Jt=\pi/4$, placing the system in a partially entangled regime locally equivalent to the $\sqrt{iSWAP}$ class, both classical equilibria persist. Upon reaching maximum entanglement at $Jt=\pi/2$, which is locally equivalent to the $iSWAP$ gate, the defection equilibrium ($D,D$) is eliminated, leaving ($C,C$) as the sole Nash equilibrium. This shows that strong entanglement resolves the coordination problem by removing the multiplicity of the equilibrium and ensuring cooperation as the only rational strategy.\\
The analyzes of the three games are summarized in Table~\ref{tab:quantum_games}. 
\begin{table*}[htbp]
\caption{Comparison of Nash equilibria in classical and quantum versions of two-player games.}
\centering
\footnotesize
\renewcommand{\arraystretch}{1.3}
\setlength{\tabcolsep}{6pt}
\begin{tabular}{|c|c|c|c|c|}
\hline
\multirow{2}{*}{Game type} 
& \multicolumn{3}{c|}{Quantum version} & Classical version \\ \hhline{|~----|}
& $c_1=c_2=Jt$ & Operator & Nash equilibrium & Classical NE \\ \hline
\multirow{3}{*}{Prisoner's dilemma}
& $0$ & $Identity$ & ($D,D$) & \multirow{3}{*}{($D,D$)} \\ 
& $\pi/4$ & $\sqrt{iSWAP}$ & ($C,C$) & \\
& $\pi/2$ & $iSWAP$ & ($C,C$) & \\ \hline
\multirow{3}{*}{Game of chicken}
& $0$ & $Identity$ & ($C,D$), ($D,C$) & \multirow{3}{*}{($C,D$), ($D,C$)} \\ 
& $\pi/4$ & $\sqrt{iSWAP}$ & $(C,D), (D,C), (C,C)$ & \\ 
& $\pi/2$ & $iSWAP$ & ($C,C$) & \\ \hline
\multirow{3}{*}{Stag hunt}
& $0$ & $Identity$ & ($C,C$), ($D,D$) & \multirow{3}{*}{($C,C$), ($D,D$)} \\ 
& $\pi/4$ & $\sqrt{iSWAP}$ & ($C,C$), ($D,D$) & \\
& $\pi/2$ & $iSWAP$ & ($C,C$) & \\ \hline
\end{tabular}
\label{tab:quantum_games}
\end{table*}
\section{Discussion}\label{sec5}
The detailed analyzes presented in the preceding sections establish how quantization via a Hamiltonian-driven entangling evolution modifies the equilibrium structure of standard two-player games. One key outcome of this study is that entanglement acts as a resource that modifies the equilibrium. In simple terms, it reshapes equilibrium behavior with its influence governed by the interaction-time product, $Jt$. In contrast to the conventional quantum game formulations that postulate a fixed entangling gate, the present approach derives strategic quantization directly from the time evolution generated by the transverse-field Ising hamiltonian. The transition to cooperative equilibria occurs consistently within the perfect-entangler region $\pi/4 \le Jt \le \pi/2$, locally equivalent to the $iSWAP$ family of gates along the $QA_2$ edge of the Weyl chamber. Accordingly, the strategic behavior acquires a clear physical interpretation rooted in Hamiltonian parameters. Also, the degree of entanglement is quantified using the concurrence, which varies monotonically with the interaction-time parameter. As a result, the effects become more pronounced as the entanglement increases, reaching a maximum in the perfect-entangler regime.\\
Across all three games considered, namely Prisoner’s Dilemma, Game of Chicken, and Stag Hunt, the classical equilibrium structure is recovered in the non-entangled limit $Jt=0$, confirming the consistency with classical game theory. As $Jt$ increases and evolution enters the entangling regime, qualitative changes in strategic behavior emerge.\\
For games characterized by social dilemmas, such as the Prisoner’s Dilemma, the introduction of entanglement resolves the conflict between individual rationality and collective welfare. The increasing entanglement transforms the dominant non-cooperative equilibrium into a Pareto-optimal cooperative outcome. In contrast, games involving anti-coordination, such as the Game of Chicken, display a suppression of asymmetric equilibria as entanglement increases, with strongly entangled regimes favoring symmetric and fair outcomes. Coordination games such as Stag Hunt exhibit yet another behavior: while multiple equilibria persist at low entanglement, increasing $Jt$ biases equilibrium selection toward the payoff-dominant cooperative state.\\
A unifying feature in all cases is that the increase in entanglement, controlled by $Jt$, systematically reshapes the Nash equilibrium structure. As the system moves from the non-entangled to the entangled regime, defection-dominated or asymmetric equilibria are progressively suppressed. At the same time,  cooperative outcomes become stable, that is, pareto optimal. Importantly, these transformations occur without modifying the underlying payoff matrices or relaxing the rationality assumptions. This suggests that entanglement itself can function as a strategic resource. This behavior is consistent across qualitatively distinct classes of games, highlighting the robustness of quantum correlations in the reshaping of equilibrium structures.\\
From an implementation point of view, the control parameter $Jt$ has direct experimental relevance. In a given physical realization such as coupled spins, superconducting qubits, or cold-atom lattices, the interaction strength 
$J$ is typically fixed by system-specific properties, while the evolution time 
$t$ can be externally tuned. Consequently, stronger interactions allow the system to reach a desired entangling operation over shorter time scales, whereas weaker couplings require a longer evolution to generate the same nonlocal effect. This trade-off highlights how strategic outcomes can be engineered either by modifying coupling strengths or by adjusting interaction times, making the framework adaptable to diverse experimental platforms. This Hamiltonian-based viewpoint thus unifies quantum game theory with concepts from quantum many-body physics and opens the possibility of studying strategic behavior using observables and tools familiar from condensed-matter systems.

\section*{Data Availability}
Data sharing is not applicable to this article, as no datasets were generated or analyzed.

\section*{Conflict of Interest}
The authors declare no conflict of interest.

\section*{Author Contributions}
All authors contributed equally to this work.

\bibliography{sn-bibliography}
\end{document}